\documentclass[prl,twocolumn,amsmath,amssymb,showpacs]{revtex4-1}
\usepackage{amsmath,amssymb,amscd}
\usepackage{hyperref}
\usepackage{enumerate}
\usepackage{graphicx}
\usepackage{mathbbol}
\usepackage{amsfonts}
\usepackage{natbib}
\usepackage{color}
\usepackage{setspace}
\begin{document}

\title{Universal Order and Gap Statistics of Critical Branching Brownian Motion}

\author{Kabir Ramola}
\email[]{kabir.ramola@u-psud.fr}
\affiliation{CNRS, LPTMS, Univ. Paris-Sud, 91405 Orsay Cedex, France}

\author{Satya N. Majumdar}
\email[]{majumdar@lptms.u-psud.fr}
\affiliation{CNRS, LPTMS, Univ. Paris-Sud, 91405 Orsay Cedex, France}

\author{Gr\'egory Schehr}
\email[]{gregory.schehr@lptms.u-psud.fr}
\affiliation{CNRS, LPTMS, Univ. Paris-Sud, 91405 Orsay Cedex, France}
\begin{abstract}
We study the order statistics of one dimensional branching Brownian motion in which 
particles either diffuse (with diffusion constant $D$), die (with rate $d$) or split 
into two particles (with rate $b$). At the critical point $b=d$ which we focus on, we show that, 
at large time $t$, the particles are collectively bunched together. 
We find indeed that there are two length scales in the system: (i) the diffusive length scale
$\sim \sqrt{Dt}$ which controls the collective fluctuations of the whole bunch and (ii) 
the length scale of the gap between the bunched particles $\sim \sqrt{D/b}$. 
We compute the probability distribution function $P(g_k,t|n)$ of the $k$th gap $g_k = x_k - x_{k+1}$ 
between the $k$th and $(k+1)$th particles given that the system contains exactly $n>k$ particles at 
time $t$. We show that at large $t$, it converges to a stationary distribution $P(g_k,t\to \infty|n) = p(g_k|n)$
with an algebraic tail $p(g_k|n) \sim 8(D/b) g_k^{-3}$, 
for $g_k \gg 1$, independent of $k$ and $n$. We verify our predictions with 
Monte Carlo simulations. 
\end{abstract}
\date{\today}
\maketitle

The statistics of the global maximum of a set of random variables
finds applications in several fields including physics, engineering, finance and geology \cite{gumbel}
and the study of such extreme value statistics (EVS) has been growing in prominence
in recent years \cite{katz,embrecht,bouchaud_mezard,dean_majumdar,
monthus,gutenburg}. In many real world examples where EVS is important,
the maximum is not independent of the rest of the set and there are
strong correlations between near-extreme values.
Examples can be found in meteorology where extreme temperatures are usually part of 
a heat or cold wave \cite{robinson} and in earthquakes and financial crashes where 
extreme fluctuations are accompanied by foreshocks and aftershocks 
\cite{omori,utsu,lillo,peterson}.
Near-extreme statistics also play a vital role in the physics of disordered systems where 
energy levels near the ground state become important at low but finite temperature 
\cite{bouchaud_mezard}.
In this context, the distribution of the $k$th maximum $x_k$ of an ordered set $\{x_1 > x_2 > x_3 ...\}$
(order statistics \cite{order_book}) and the gap between successive maxima $g_k = x_k - x_{k+1}$  
provides valuable information  
about the statistics near the extreme value.
Such near-extreme distributions have recently been of interest in statistics \cite{pakes} 
and physics \cite{sabhapandit_majumdar,schehr_majumdar,mounaix,perret}.
Although the order and gap statistics of independent 
identically distributed (i.i.d.) variables are fully understood~\cite{order_book},
very few exact analytical results exist for strongly correlated random variables.
In this context, random walks and Brownian motion offer a fertile arena 
where near-extreme distributions for correlated variables can be computed analytically 
\cite{racz,schehr_majumdar,mounaix,perret}.

\begin{figure}
\includegraphics[width=\linewidth]{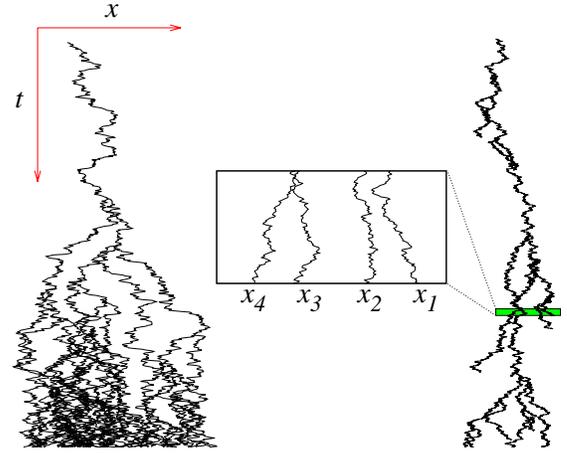}
\caption{A realization of the dynamics of branching Brownian motion with death (left) in 
the supercritical regime ($b > d$) and (right) the critical regime ($b = d$). 
The particles are numbered sequentially from right to left
as shown in the inset.}
\label{Fig1}
\end{figure}

Another interesting system where order statistics plays an important role is the 
branching Brownian motion (BBM). In BBM, a single particle starts initially
at the origin. Subsequently, in a small time interval $dt$, the particle 
splits into two independent offsprings with probability $b\, dt$, dies
with probability $d\, dt$ and with the remaining probability $(1- (b+d)\,dt)$
it diffuses with diffusion constant $D$.
A typical realization of this process is shown in Fig. \ref{Fig1}. 
BBM is a prototypical model of evolution, but has also
been extensively used as a simple model for reaction-diffusion systems, 
disordered systems, nuclear reactions, cosmic ray showers, 
epidemic spreads amongst others 
\cite{brunet_derrida_epl,brunet_derrida_jstatphys,mezard,derrida_spohn,demassi, 
takayasu,harris,golding,fisher, 
sawyer,bailey,mckean,bramson,majumdar_pnas,derrida_brunet_simon}. In one dimension, the 
position of the existing particles at time $t$ constitute a set of strongly correlated 
variables that are 
naturally ordered according to their positions on the line with $x_1(t) > x_2(t) > x_3(t) 
\hdots$. The particles are labelled sequentially from right to left as shown in Fig. 
\ref{Fig1}. One dimensional BBM then provides a natural setting to study
the order and the gap statistics for strongly correlated variables.


The number of particles $n(t)$ present at time $t$ in this process is a random variable 
with different behavior depending on the relative magnitude of the rates of birth $b$ and 
death $d$. When $b < d$ ({\em{subcritical}} phase), the process dies eventually and on an 
average there are no particles at large times. In contrast, for $b > d$ 
({\em{supercritical}} phase), the process is explosive and the average number of 
particles grows exponentially with time. In the borderline $b=d$ ({\em{critical}}) case, 
the probability $P(n,t)$ of having $n$ particles at time $t$, starting with a single
particle initially, has a
well known expression~\cite{feller}
(a simple derivation is provided in~\cite{supplementary})
\begin{eqnarray}
P(0,t)=\frac{bt}{1+bt} \;, \; P(n\ge1,t)=\frac{(bt)^{n-1}}{(1+bt)^{n+1}} \;.
\label{particle_probabilities}
\end{eqnarray}
The probability that there are no particles tends to $1$
as $1-1/(bt)$ while the probability that there are $n \ge 1$ particles
tends to $0$ as $1/(bt)^2$. The average number of particles is independent of time
with $\langle n(t) \rangle =\sum_{n=1}^{\infty}n P(n,t)=1$.
There are thus strong fluctuations at the critical point which causes most of
the realizations of this process to have no particles at large times.

In the supercritical phase, in particular for $d=0$, the statistics of the $k$-th 
maximum $x_k(t)$
has been studied extensively in mathematics and physics literature with
direct relevance to polymer~\cite{derrida_spohn} and spin-glass physics~\cite{mezard}. 
For example, the
first maximum $x_1(t)\sim v t$ typically increases linearly with $t$
and its cumulative distribution satisfies a nonlinear Fisher-Kolmogorov-Petrovky-Piscounov
equation~\cite{fisher,kpp} with a traveling front solution with velocity 
$v$~\cite{mckean,bramson}. The statistics of this first maximum, in the
supercritical phase, also
appears in numerous other applications in mathematics~\cite{lalley_sellke,arguin} and  
physics~\cite{brunet_derrida_epl,brunet_derrida_jstatphys,majumdar_pnas}. 
More recently, the statistics of the gaps between successive maxima have
also been studied in the supercritical phase~
\cite{brunet_derrida_epl,brunet_derrida_jstatphys} and
the average gap between the $k$-th and $(k+1)$-th maximum was shown to tend to a 
$k$-dependent constant, independent of time $t$, at large $t$. The stationary probability distribution function (PDF) of the first gap was also computed numerically and an analytical argument was given to explain its exponential tail~\cite{brunet_derrida_epl,brunet_derrida_jstatphys}.
However, an exact analytical computation of the stationary PDFs of 
these gaps in the supercritical phase still remains an open problem.

Much less is known about the order statistics at the critical point ($b = d$)
which is relevant to several systems
including population dynamics, epidemics spread, nuclear reactions
etc.~\cite{majumdar_pnas,lalley,sagitov,aldous}. In this Letter,
we show that, in contrast to the supercritical case, 
the order and the gap statistics can be computed exactly 
for the critical case $b=d$.
In the critical case where $\langle n(t)\rangle=1$
at all times, to make sense of the gaps between particles, it is necessary to work in the 
fixed particle number sector, i.e.,
condition the process to have
exactly $n(t)=n$ particles at time $t$, with their ordered positions 
denoted by $x_1(t) > \hdots x_k(t) \hdots > x_n(t)$. We show that
a typical trajectory of the critical process is characterized by two
length scales at late times: (i) each particle   
$\langle |x_k(t)| \rangle \sim \sqrt{4Dt/\pi}$ for all 
$1 \le k \le n$, implying an effective bunching of the particles into
a single cluster that diffuses as a whole and (ii) within this bunch, the gap
$g_k(t)= x_k(t) - x_{k+1}(t)$ between
successive particles tends to a time-independent random variable of $\sim O(1)$.
We compute analytically the PDF of this gap
(conditioned to
be in the fixed $n$-particle sector) and show that it becomes stationary at late times
$P(g_k=z,t\to \infty|n)\to p(z|n)$ independent of $k$. Moreover, quite {\em remarkably},
$p(z|n)$ has an {\em universal} algebraic tail, 
$p(z|n)\sim 8(D/b)/{z}^3$, independent of $k$ and $n$.
 
{\it Statistics of the Maximum:} We first analyze the behavior of the rightmost particle 
at time $t$. A convenient quantity is the joint probability that there are 
$n \ge 1$ particles at
time $t$, with all of them lying to the left of $x$: $Q(n;x,t)={\rm Prob.}[n(t)=n, 
x_n(t)<x_{n-1}(t)<\ldots< x_1(t)<x]$. It evolves via a backward Fokker-Planck (BFP) 
equation which can be derived by splitting the time interval $[0,t+\Delta t]$ into
$[0,\Delta t]$ and $[\Delta t, t + \Delta t]$ and considering all events
that take place in the first small interval $[0,\Delta t]$. In this small
interval, the single particle at the origin can: i) with a probability $b \Delta t$ split 
into two
independent particles which give rise to $r$ and $n-r$ particles at the final time 
respectively;
ii) die with the probability $d \Delta t$ and therefore not contribute to the probability
at subsequent times; or iii) diffuse by a small amount $\Delta x$
with probability $1-(b +d)\Delta t$,
effectively shifting the entire process by $\Delta x$. Summing these contributions, 
taking the $\Delta t\to 0$ limit and setting $b=d$, we get~\cite{supplementary}
\begin{eqnarray}
\nonumber
\frac{\partial Q(n;x,t)}{\partial t} = 
D\frac{\partial^{2}Q(n;x,t)}{\partial x^{2}} - 2 b Q(n;x,t)\\
+ 2 b\, P(0;t)Q(n;x,t)+ b\, \sum_{r = 1}^{n-1}Q(r;x,t)Q(n-r;x,t),
\label{BFP_Qn}
\end{eqnarray}
starting from the initial condition $Q(n;x,0)= \delta_{n,1}$ for all $x>0$
and satisfying the boundary conditions: $Q(n;-\infty,t)=0$ and $Q(n;\infty,t)=P(n,t)$.
Next, we consider the conditional probability $Q(x,t|n)=Q(n;x,t)/P(n,t)$, i.e.,
the cumulative probability of the maximum given $n$ particles at time $t$.
Using (\ref{BFP_Qn}) and
the explicit expression of $P(n,t)$ in (\ref{particle_probabilities}),
we find that $Q(x,t|n)$ evolves via
\begin{eqnarray}
\nonumber
\frac{\partial Q(x,t|n)}{\partial t} +\frac{n-1}{t(1+bt)} Q(x,t|n) =
D\frac{\partial^{2}Q(x,t|n)}{\partial x^{2}}\\ +
 \frac{1}{t(1+bt)} \sum_{r = 1}^{n-1}Q(x,t|r)Q(x,t|n-r).
\label{cond.n}
\end{eqnarray}
This is a linear equation for $Q(x,t|n)$ for a given $n$ that involves, as source
terms, the solutions $Q(x,t|k)$ with $k<n$. Hence it can be solved recursively
for any $n$, starting with $n=1$. For $n=1$, one obtains an explicit 
solution~\cite{supplementary}: $Q(x,t|1)=\frac{1}{2}\,
\mathrm{erfc}\left(\frac{-x}{\sqrt{4Dt}}\right)$, 
where $\mathrm{erfc}(x)=\frac{2}{\sqrt{\pi}}\,\int_x^{\infty} e^{-u^2}\, du$ is the 
complementary error function. Consequently, the PDF of the maximum $x_1(t)$ in the 
single
particle sector, $P(x_1,t|1)= \partial_{x_1}Q(x_1,t|1)= \frac{1}{\sqrt{4 \pi D t}} \exp 
\left(-\frac{x_1^2}{4 D t} \right)$, is a simple Gaussian. The particle thus exhibits 
free diffusion, implying that the 
effect of branching exactly
cancels the effect of death. For later purpose, we note that $P(1;x,t)=\partial_x 
Q(1;x,t)=P(1,t) \partial_x Q(x,t|1)$, i.e. the probability density of having one 
particle at 
position $x$ at time $t$, reads 
\begin{equation}
P(1;x,t)= \frac{1}{(1+bt)^2}\, \frac{1}{\sqrt{4\pi Dt}}\, e^{-x^2/{4Dt}}.
\label{p1}
\end{equation}
Finally, 
feeding the one particle solution $Q(x,t|1)$ into (\ref{cond.n}) for 
$n=2$, one can also obtain $Q(x,t|2)$ (see~\cite{supplementary}) and recursively 
$Q(x,t|n)$ for higher $n$. 

For general $n>1$, one
can estimate easily the late time asymptotic solution.
Since $Q(x,t|n)$ is bounded as $0 < Q(x,t|n) < 1$, 
Eq.~ (\ref{cond.n}) reduces, for large $t$,
to a simple diffusion equation which does not contain $n$ explicitly, implying 
$Q(x,t|n)\sim Q(x,t|1)$.
Hence, the PDF of the maximum for any $n\ge 1$
particle sector behaves as $P(x_1,t|n) \approx \frac{1}{\sqrt{4 \pi D t}}
\exp \left(-\frac{x_1^2}{4 D t} \right)$ for large $t$.
By symmetry, the minimum $x_n$ is also governed by the same distribution.   
This illustrates an important feature of BBM at criticality:
{\it the maximum and minimum of $n$ particles both behave as a 
free diffusing particle at large $t$}.
The rest of the particles are confined between these two extreme values 
($x_1(t) > \hdots x_k(t) \hdots > x_n(t)$) and hence also behave diffusively,  
$\langle |x_k| \rangle \sim \sqrt{4Dt/\pi} $, independent of $k$ and $n$ 
for large $t$, leading to 
the bunching of the particles. The gap between the particles
$g_k(t)= x_k(t)-x_{k+1}(t)$ thus probes the sub-leading large $t$ behavior of the
particle positions $x_k(t)$, which we consider next.

\begin{figure}
\includegraphics[width=\linewidth]{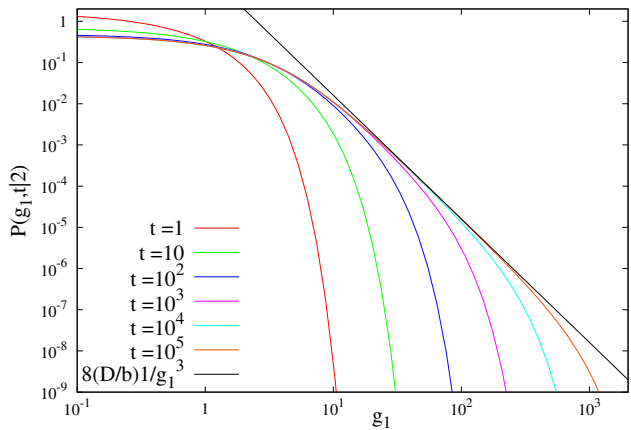}
\caption{Exact gap PDF in the two particle sector (Eq. \ref{gap_distribution}) at different times, 
showing the approach to the stationary behavior at large times. The solid line indicates the expected power 
law decay for $t \to \infty$.
Here $D = 1$ and $b ={1}/{2}$.}
\label{2part_approach}
\end{figure}

{\it Gap Statistics:} We start with the first gap $g_1(t)=x_1(t)-x_2(t)$ between
the rightmost and the preceding particle in the particle number $n\ge 2$ sector.  
To probe this gap, it is convenient to study
the joint PDF $P(n;x_1, x_2,t)$ that there are $n$ particles at 
time $t$ with the first particle at position $x_1$ and the second at position $x_2<x_1$. 
We first analyze the simplest case $n=2$ and argue later that the behavior of $g_1$
in this $n=2$ sector is actually quite generic and holds for higher $n$ as well.
Using a similar BFP approach outlined before, we find the following evolution
equation (for detailed derivation see ~\cite{supplementary})
\begin{eqnarray}
\nonumber 
\frac{\partial P(2;x_{1},x_{2},t)}{\partial t} = D\left( \frac{\partial}{\partial x_1}
+ \frac{\partial}{\partial x_2}\right)^2 
P(2;x_{1},x_{2},t)\\
-\frac{2b}{1+bt}P(2;x_{1},x_{2},t)+ 2 b P(1;x_1,t) P(1;x_2,t)
\label{2part_FP}
\end{eqnarray}
where $P(1;x,t)$ is given in (\ref{p1}). This linear equation for 
$P(2;x_1,x_2,t)$ can be solved explicitly~\cite{supplementary}. Consequently, 
the conditional probability $P(x_1,x_2,t|2)= P(2;x_1,x_2,t)/P(2,t)$
(with $P(2,t)= bt/(1+bt)^3$ given in (\ref{particle_probabilities})),
denoting the joint PDF of $x_1$ and $x_2$ given $n=2$ particles, can
also be obtained explicitly. 
The solution
is best expressed in terms of the variables, $s=(x_1+x_2)/2$ (center of mass)
and $g_1=x_1-x_2$ (gap): $P(x_1,x_2,t|2)\to P(s,g_1,t|2)$ and reads~\cite{supplementary}  
\begin{equation}
{P}(s,g_1,t|2) = \left(\frac{1 + b t}{2 \pi D t}\right) 
\int_{0}^{t} \frac{dt'}{(1+ b t')^2} \frac{e^{- \frac{g_1^2}{8 D t'}
- \frac{s^2}{2D(2 t- t')}}}{\sqrt{t'(2t-t')}}.
\label{2part_distribution}
\end{equation}
The marginal PDF of the centre of mass $P(s,t|2)=\int_{0}^{\infty} P(s,g_1,t|2) 
dg_1$ is easily obtained by integrating over
the gap $g_1$ and for large $t$,
$P(s,t|2) \sim \frac{1}{\sqrt{4 \pi D t}} \exp \left(-\frac{s^2}{4 D t} \right)$,
as expected from the free diffusive behavior of the clustered particles. 
Similarly, by integrating over $s$ we obtain the marginal PDF of the gap at any $t$
\begin{equation}
P(g_1,t|2) = \left(\frac{1 + b t}{b t}\right) \int_{0}^{t} \frac{b dt'}{(1+ b t')^2}
\frac{\exp(- \frac{g_1^2}{8 D t'})}{\sqrt{2 \pi D t'}}.
\label{gap_distribution}
\end{equation}
At large times $P(g_1,t|2)$ converges to a stationary distribution 
$P(g_1,t \to \infty|2) = p(g_1|2)$ 
(Fig. \ref{2part_approach}), which can be computed explicitly. It can be expressed as 
$p(g_1|2) = (4\sqrt{D/b})^{-1} f[g_1/(4\sqrt{D/b})]$ with
\begin{eqnarray}
\label{scaling_f}
f(x) = -4x + \sqrt{2\pi} \, e^{2x^2}(1+4x^2)\,{\rm erfc}(\sqrt{2} \,x) \;.
\end{eqnarray}
This distribution (\ref{scaling_f}) has a very interesting relation to the 
PDF of the (scaled) $k$-th gap between extreme points visited by a single random walker 
found in Ref. \cite{schehr_majumdar} [the scaling function found there (see Eq. (1) of \cite{schehr_majumdar}) 
is exactly $-f'(x)/\sqrt{2 \pi}$]. It behaves asymptotically as  
\begin{eqnarray}
p(g_1|2) \sim 
\begin{cases}
\sqrt \frac{\pi b}{8 D} \;, \;  g_1 \to 0,\\
\left(\frac{8 D}{b} \right) {g_1^{-3}} \;, \; g_1 \to \infty \;.
\end{cases}
\label{large_g_behaviour}
\end{eqnarray}
This function $p(g_1|2)$ describes the typical fluctuations of the gap $g_1$, which are of order $\sqrt{D/b}$. 
However, because of the algebraic tail, only the first moment of the gap is dominated by the typical fluctuations, 
$\langle g_1 \rangle = \sqrt{{2 \pi D}/{b}}$. The higher moments instead get contributions from the time dependent
far tail of the PDF in (\ref{gap_distribution}): $\langle g_1^2 \rangle \sim \ln(t)$ and 
$\langle g_1^m \rangle \sim t^{\frac{m}{2}-1}$ for $m >2$.
In Fig. \ref{2part_approach}, we plot $P(g_1,t|2)$ at different times showing the
approach to the stationary distribution with a power law tail at large times.

\begin{figure}[hh]
\includegraphics[width=\linewidth]{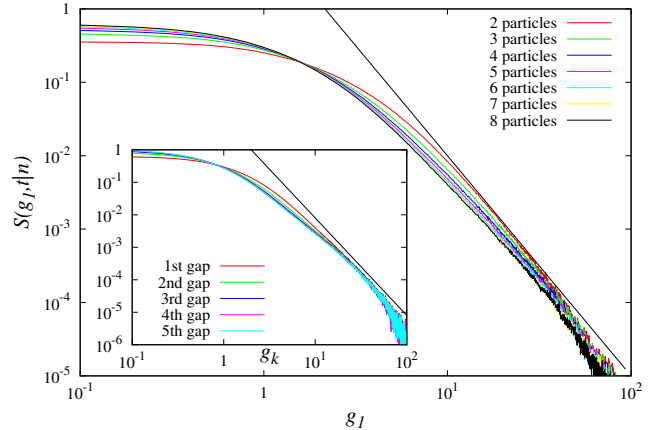}
\caption{Time-integrated PDF for the first gap $g_1 = x_1 - x_2$ in different particle sectors computed from 
Monte Carlo simulations. {\bf Inset}: Time-integrated PDF for the $k$-th gap $g_k = x_k - x_{k+1}$ in the $10$-particle sector, 
showing the approach to the same asymptotic value. 
The lines have a slope of $-3$. Here $D= 1, b = {1}/{2}$, and $t =10^4$.}
\label{Npartfit}
\end{figure}

The computation for the first gap $g_1$ for $n=2$ outlined above can be
generalized to the $n>2$ sector.
Once again using the BFP approach, we find that the joint PDF $P(n;x_{1},x_{2},t)$ obeys
\begin{eqnarray}
\nonumber
&&\frac{\partial P(n;x_{1},x_{2},t)}{\partial t} = D\left( \frac{\partial}{\partial x_1} + 
\frac{\partial}{\partial x_2}\right)^2 P(n;x_{1},x_{2},t)\\
&&~~~~~~~~~~~-\frac{2b}{1+bt}P(n;x_{1},x_{2},t)+ b \mathcal{S}(n;x_1,x_2,t).
\label{n_particle_diffusion}
\end{eqnarray}
Here $\mathcal{S}(n;x_1,x_2,t)$ is a source term that arises from the branching at the 
first time step. It can be computed explicitly in terms of spatial integrals involving 
$P(k; x_1, x_2, t)$ with $k < n$ -- the resulting expression being however a bit cumbersome \cite{supplementary}. 
%
%
%
%
However Eq. (\ref{n_particle_diffusion}) can still be solved recursively to obtain the exact distribution of the first gap $g_1 = x_1 - x_2$ in the $n$ particle sector. 
We have solved these equations exactly up to $n=4$ \cite{supplementary}. 
These computations are quite instructive as they allow us to analyze Eq. (\ref{n_particle_diffusion}) in the large $t$ 
and large gap $g_1$ limit for generic $n$ as follows. 
The solution of~(\ref{n_particle_diffusion}) is a linear combination of solutions arising from individual terms present 
in the source function ${\cal S}$. From this one can show that the PDF of the first gap in the $n$-particle
sector converges to a stationary distribution $P(g_1,t\to \infty|n)=p(g_1|n)$. 
While the full PDF $p(g_1|n)$ depends on $n$ (see also Fig. \ref{Npartfit}), its tail is universal.
This follows from the fact that the leading contribution to ${\cal S}$ in (\ref{n_particle_diffusion})
when the gap $g_1 = x_1-x_2 \gg 1$ is large 
%
tends to $2 b P(1;x_1,t) P(1;x_2,t)$ at large $t$ \cite{supplementary}. 
This is precisely the source term for the two-particle case 
analyzed in Eq. (\ref{2part_FP}). One can show that all other terms in ${\cal S}$ involve a larger gap between 
particles generated by the same offspring walk and are thus suppressed by a 
factor $\int_{g_1}^{\infty} p(g'|k) dg'$, $k<n$ \cite{supplementary}. 
Therefore, when $g_1 \to \infty$ the tail of the PDF of the first gap in the $n$ 
particle sector converges to that of the two-particle case, 
$p(g_1|n) \sim \left(\frac{8 D}{b} \right) {g_1^{-3}}$, for all~$n$. 

A similar analysis yields the asymptotic behavior of the $k$-th gap $g_k(t) = x_{k}(t) - x_{k+1}(t)$. 
In this case, we study $P(n;x_k,x_{k+1},t)$, the joint PDF that 
there are $n$ particles at time $t$ with the $k$-th particle at position $x_k$ and 
the $(k+1)$-th particle at position $x_{k+1}$. This PDF once again satisfies a diffusion equation 
with a source term similar to~(\ref{n_particle_diffusion}), from which we can show that the PDF of the $k$th gap
reaches a stationary distribution $P(g_k,t\to \infty|n)=p(g_k|n)$. In the large gap limit, 
the dominant term in the source function is the one in which the 
first $k$ particles belong to one of the offsprings generated at the first time step, 
and the subsequent $n-k$ particles belong to the other.
This term tends to $2 b P(1;x_k,t) P(1;x_{k+1},t)$ at large $t$, as it involves the minimum of
the first process being at $x_k$ and the maximum of the other process being at $x_{k+1}$. 
As noticed before for $g_1$, all other terms involve a large gap between particles generated by the same offspring process
and are hence suppressed. This in turn leads to the large gap stationary behavior 
$p(g_k|n) \sim \left(\frac{8 D}{b} \right)  g_k^{-3}$ for all $k$ and $n$. 

{\it Monte Carlo Simulations:} We have directly simulated the critical 
BBM process and we have computed the PDFs of the gap. 
To obtain better statistics we compute the time-integrated PDF
$S(g_k,t|n) = \frac{1}{t}\int_{0}^{t} P(g_k,t'|n) dt'$, which has the same stationary 
behavior as $P(g_k,t|n)$, $S(g_k,t\to \infty|n) = p(g_k|n)$. In Fig. \ref{Npartfit} we plot $S(g_1,t|n)$, 
corresponding to the first gap, 
for different values of $n = 1, \cdots, 8$ and $t=10^4$. The different curves show an approach 
to the same asymptotic, large $g_1$, behavior (note that the approach to the stationary state gets slower as $n$ increases). 
In the inset of Fig. \ref{Npartfit} we show a plot of $S(g_k,t|n)$ for $n=10$ and 
$t = 10^4$ for different values of $k=1, \cdots, 5$. This also 
shows a convergence to the same large $g_k$ behavior $\sim  \left(\frac{8 D}{b} \right)  g_k^{-3}$. 
Numerical results for short times (up to $n=4$), not shown here \cite{supplementary},
show a perfect agreement with the solution of Eq. (\ref{n_particle_diffusion}).

{\it Conclusion}: We have obtained exact results for the order statistics of critical BBM. We showed that  
the statistics of the near extreme points displays a quite rich behavior characterized by a 
stationary gap distribution with a universal algebraic tail. This presents a physically relevant instance
of strongly correlated random variables for which order statistics can be solved exactly.  
It will be interesting to extend the BFP method developed here to compute exactly 
the gap statistics 
in the supercritical case.

\acknowledgments{} KR acknowledges helpful discussions with Shamik Gupta. 
SNM and GS acknowledge support by ANR grant 2011-BS04-013-01 WALKMAT and in part by the 
Indo-French Centre for the Promotion of Advanced Research under Project 4604-3. 
GS acknowledges support from Labex-PALM (Project Randmat). 



\end{document}